\documentclass[11pt]{article}
\usepackage{citesort}
\usepackage{gslist}
\usepackage{gsnotat}
\usepackage{ftp97}

\evensidemargin\oddsidemargin
\begin{document}
\title{Testing for Renamability to Classes of Clause Sets%
  \thanks{Cite as ``Albert Brandl, Christian G.~Ferm\"uller,
    Gernot Salzer: Testing for Renamability to Classes of Clause Sets.
    In \emph{Proc. Int.\ Workshop on First-Order Theorem Proving (FTP'97)},
    Maria Paola Bonacina and Ulrich Fuhrbach, editors.
    RISC-Linz Report Series No.\ 97-50, pages 34--39.
    Johannes Kepler Universit\"at, Linz, Austria, 1997.''}\ %
  \thanks{Supported by FWF grant P11624-MAT.}%
}
\author{\begin{tabular}{c@{\hspace{10mm}}c@{\hspace{10mm}}c}
   Albert Brandl & Christian G.~Ferm\"uller & Gernot Salzer%
   \thanks{\texttt{gernot.salzer@tuwien.ac.at}}\\
   \multicolumn{3}{c}{Technische Universit\"at Wien, Austria}
   \end{tabular}%
}
\date{}
\maketitle
\thispagestyle{plain}

\begin{abstract}
   This paper investigates the problem of testing clause sets
   for membership in classes known from literature.
   In particular, we are interested in classes defined via renaming:
   Is it possible to rename the predicates in a way such that
   positive and negative literals satisfy certain conditions?
   We show that for classes like \emph{Horn} or
   \emph{\OCCiN~\cite{FLTZ93}} the existence of such
   renamings can be decided in polynomial time, whereas
   the same problem is NP-complete for class \emph{\PVD~\cite{FLTZ93}}.
   The decision procedures are based on hyper-resolution; if a renaming
   exists, it can be extracted from the final saturated clause set.
\end{abstract}

\section{Introduction}

Within the last decade the classical decision problem for predicate logic
was revisited in the frame of clause logic. Instead of deciding the validity
of first order formulas in prenex normal form---usually without function
symbols---the aim was now to decide the satisfiability of clause sets with
a rich functional structure. Many classes of clause sets---with and without
equality---were shown to be efficiently decidable using refinements 
of resolution, paramodulation and superposition~\cite{%
Joyner76ACM,FLTZ93,FermSalz93LPAR,BachmairGanzingerWaldmann93KGC},
subsuming most of the classical results. For some classes it was
possible to go even further: the output of the decision procedure can be
used to derive finite representations of models~\cite{%
FermuellerLeitsch93,FermLeit98JIGPL}.

When implementing decision procedures one encounters an additional
problem: before actually submitting a clause set to the decision procedure for
a class~$\gccs$, one has to check that the clause set satisfies all criteria
characterizing~$\gccs$. Applying a decision procedure to clause sets outside of
its scope might yield uninterpretable results or lead to non-termination.
For many classes this check is quite simple as
the criteria can be tested locally one literal or one clause at a time.
However, some classes involve global conditions like renaming;%
\footnote{\emph{Renaming} a predicate symbol~$P$ here means 
mirroring the polarity of~$P$ throughout the whole clause set
by replacing all occurrences of $\neg P(\cdots)$ by $P(\cdots)$ and
all occurrences of $P(\cdots)$ by $\neg P(\cdots)$.
The renamed clause set is similar to the original one in many respects;
e.g., they are equivalent with respect to satisfiability.}
a na\"{\i}ve algorithm would consider all possible renamings, whose number
is exponential in the number of different predicate symbols.

\begin{example}\label{xa:intro}
   To find out whether a given clause set is Horn
   one has simply to check that the number of positive literals per clause
   is zero or one. Obviously this test is linear in the size of the clauses.
   However, it is more complex to check whether a clause set
   can be \emph{made} Horn by renaming some of its predicate symbols.
   E.g., the clause set
   \(\set{\set{P(x),Q(x)},\,\set{\neg P(y),\,\neg Q(y)}}\)
   is not Horn, but can be made Horn by renaming $P$ (or alternatively~$Q$),
   yielding the set \(\set{\set{\neg P(x),Q(x)},\,\set{P(y),\,\neg Q(y)}}\).
\end{example}

In this paper we investigate three instances of the renamability problem.
After fixing some notions and notations in section~\ref{sec:notat},
we describe in section~\ref{sec:Horn} a technique---introduced by Harry R.~Lewis
in~\cite{Lewis78}---for testing renamability to Horn.%
\footnote{We thank one of the referees 
for pointing out to us that there exists an impressive body of literature
on renamability to Horn. In particular, we learned that we re-discovered 
Lewis' proof. (We apologize for not having done our homework.)}
In~\ref{sec:OCC1N} we show that the renamability problem for
\emph{\OCCiN} can be decided by a related method
in polynomial time. 
Section~\ref{sec:PVD} proves that the renamability problem for~\emph{\PVD} is
NP-complete. The final section discusses how suitable renamings
can be extracted from the information generated by the decision procedures.

\section{Basic Notions}\label{sec:notat}

For basic notions like clause, literal, etc.\ we refer the reader
to textbooks like~\cite{ChangLee}.

The \emph{dual} of a literal~$\gl$ is denoted by $\dual\gl$ and
is defined by $\dual{P(\cdots)}=\neg P(\cdots)$ and
$\dual{(\neg P(\cdots))}=P(\cdots)$.
The \emph{propositional skeleton} of~$\gl$ is denoted by~$\skel(\gl)$
and is defined as $\skel(\optneg P(\cdots))={\optneg P}$, where the second
occurrence of~$P$ is interpreted as propositional variable.
$\skel$~is extended to clauses and sets of clauses in the obvious way.
The set of
variables occurring in a term, atom, literal or clause~$E$ is denoted by
$\var(E)$. By $\occ(v,E)$ we denote the number of occurrences of variable~$v$
in~$E$.
For a clause~$\gcl$, the subset of its positive literals is denoted by~$\pc\gcl$,
the subset of its negative literals by~$\nc\gcl$. 

A {\em renaming} is a set of predicate symbols. The application of a
renaming $\gr$ to a literal~$\gl$, denoted by $\gr(\gl)$, is $\dual\gl$
if the predicate symbol of~$\gl$ occurs in $\gr$, and $\gl$ otherwise.
The result of renaming a clause $\gcl= \set{\gl_1,\ldots,\gl_n}$
by $\gr$ is $\gr(\gcl) = \set{\gr(\gl_1),\ldots,\gr(\gl_n)}$.
Similarly, a clause set $\gcs=\set{\gcl_1,\ldots,\gcl_n}$
is renamed to $\gr(\gcs) = \set{\gr(\gcl_1),\ldots,\gr(\gcl_n)}$.
For every model~$\M$ of a propositional clause set,
$\gr_\M$ denotes the set of propositional variables that are true in~$\M$.

The \emph{depth}~$\td(t)$ of a term~$t$ is defined as $\td(t)=0$ if
$t$ is a constant or variable, and as
$\td(f(t_1,\ldots,t_n))=1+\max\Set{\td(t_i)}{1{\leq}i{\leq}n}$
for a functional term. For an atom or literal we define
$\td(\optneg P(t_1,\ldots,t_n))=\max\Set{\td(t_i)}{1{\leq}i{\leq}n}$.
If $\gcl$ is a clause then $\td(\gcl)$ is an
abbreviation for $\max\Set{\td(\gl)}{\gl\in\gcl}$.
The \emph{maximal depth of occurrence}~$\tmax(\gv,t)$ of a variable~$\gv$
in a term~$t$ is defined by $\tmax(\gv,\gv)=0$ and
$\tmax(\gv,f(t_1,\ldots,t_n))=
  1+\max\Set{\tmax(\gv,t_i)}{\gv\in\var(t_i),\,1{\leq}i{\leq}n}$.
Analogously, we define
the \emph{minimal depth of occurrence}~$\tmin(\gv,t)$. 
These definitions are extended to atoms and literals in the obvious way.

\section{Testing for Renamability to Horn}\label{sec:Horn}

\begin{definition}
   A clause set, $\gcs$, is Horn iff each of its clauses contains
   at most one positive literal.
   $\gcs$ is renamable to Horn if there is a renaming,~$\gr$,
   such that $\gr(\gcs)$ is Horn.
\end{definition}

The importance of the Horn fragment of clause logic is well known. 
It is also well known that clause sets that are not Horn can often
be converted to Horn by systematically renaming 
the predicate symbols (see example~\ref{xa:intro}).
The fact that renamability to Horn is decidable
in polynomial, even linear time is well documented in the literature
(see e.g.~\cite{Lewis78,MannilaMehlhorn85,KleineBuening90,Hebrard94}).
However, since we feel that our proof technique
is best explained for the case of clause sets renamable to Horn we briefly
re-describe Lewis' idea~\cite{Lewis78} of converting the renamability problem
into a satisfiability problem for sets of propositional Krom clauses.
The significance of the method lies in the fact that it can be
adapted to many similar problems; see e.g.\ section~\ref{sec:OCC1N}.
We illustrate the method by an example.
\begin{example}\label{xa:Horn}
Let $\gcs$ be the set containing the clauses
\( \gcl_1 = \set{P(x), Q(x), R(x)} \),
\( \gcl_2 = \set{\neg P(y), Q(y)} \),
\( \gcl_3 = \set{\neg R(x)} \), and
\( \gcl_4 = \set{\neg P(x), \neg Q(x)} \).
Obviously $\gcl_1$ is not Horn, therefore $\gcs$ is not Horn.
Of course $\gcl_1$ can be
renamed to Horn, e.g.\ by applying the renaming $\gr = \set{P,Q}$. 
$\gr(\gcl_2)$ and $\gr(\gcl_3)$ are Horn, too,
but $\gr(\gcl_4)$ is not. Therefore we have to `backtrack'
and to try another candidate for a renaming to Horn.
The question is whether we can compute an
appropriate renaming $\gr$ without backtracking. 
Observe that $\gcl_1$ imposes the following restriction on~$\gr$
if $\gr(\gcl_1)$ is to be Horn:
\begin{center}
   either $P \in \gr$ or $Q \in \gr$;
   and either $P \in \gr$ or $R \in \gr$;
   and either $Q \in \gr$ or $R \in \gr$.
\end{center}
This just expresses the fact that for every pair of literals
in a Horn clause at least one of the two literals has to be negative.
Similarly, we obtain the following condition for the only pair of
literals in $\gcl_2$:
\begin{center}
   either $P \notin \gr$ or $Q \in \gr$.
\end{center}
Since $\gcl_3$ is singleton there is no corresponding condition:
all singleton clauses are Horn by definition. For $\gcl_4$ we obtain:
\begin{center}
   either $P \notin \gr$ or $Q \notin\gr$.
\end{center}
The conditions are simultaneously satisfiable. This can easily be seen
by representing them as a set of propositional Krom%
\footnote{A Krom clause is a clause containing at most two literals.}
clauses.
If we abbreviate the proposition \emph{`$P\in\gr$'} by $P$
and similarly \emph{`$P\notin\gr$'} by $\neg P$
(interpreting the predicate symbol as a propositional variable)
we obtain the clause set
\begin{eqnarray*}
   \ren(\gcs) &=&
   \set{\set{P,Q},\,
        \set{P,R},\,
        \set{Q,R},\,
        \set{\neg P,Q},\,
        \set{\neg P,\neg Q}}\enspace.
\end{eqnarray*}
A model for $\ren(\gcs)$ is given 
by setting $Q$ and $R$ to true and $P$ to false.
It corresponds to the renaming $\gr = \set{Q,R}$.
In fact, the set of models
for $\ren(\gcs)$ represent \emph{all} renamings~$\gr$
for which $\gr(\gcs)$ is Horn. 
\end{example}

For a clause $\gcl=\set{\gl_1,\ldots,\gl_n}$,
let $\ren(\gcl)$ be the set
$\Set{\set{\skel(L_i),\skel(L_j)}}{i\neq j}$
of propositional Krom clauses.
For a clause set $\gcs$, let $\ren(\gcs) = \bigcup_{\gcl\in\gcs}\ren(\gcl)$.

\begin{proposition}[Lewis \cite{Lewis78}]
For every clause set $\gcs$, $\gr_\M(\gcs)$ is Horn iff $\M$
 is a model of~$\ren(\gcs)$.
\end{proposition}

The satisfiability of propositional Krom clause sets can be tested in polynomial
time (e.g.\ by computing all resolvents). Moreover, $\ren(\gcs)$ is of at most
quadratic size with respect to the size of~$\gcs$
and can easily be computed in polynomial time.
Therefore renamability to Horn can be tested in polynomial time.
More importantly, the renamings can be efficiently computed by
hyper-resolution and splitting (see section~\ref{sec:renaming}).

\section{Testing for Renamability to \OCCiN}\label{sec:OCC1N}

\begin{definition}
   A clause set,~$\gcs$, is in class~\OCCiN\ iff every clause $\gcl\in\gcs$
   satisfies the following conditions:
   \begin{labenum}{OCC}
      \item \label{cond:OCCa}
            $\occ(\gv,\pc\gcl)=1$ for all $\gv\in \var(\pc\gcl)$;
      \item \label{cond:OCCb}
            $\tmax(\gv,\pc\gcl)\leq\tmin(\gv,\nc\gcl)$
            for all $\gv\in\var(\nc\gcl)\cap\var(\pc\gcl)$.
   \end{labenum}
   $\gcs$ is renamable to~\OCCiN\ if there is a renaming,~$\gr$,
   such that $\gr(\gcs)$ is in~\OCCiN.
\end{definition}

In~\cite{FLTZ93} it is shown that the satisfiability of clause sets
in~\OCCiN\
can be decided by hyper-resolution. Therefore the class of clause sets
which are renamable to~\OCCiN\ is decidable, too: just apply the decision
procedure to the renamed clause set.

Like in the case of renamability to Horn we encode the 
class membership conditions for each candidate clause 
by propositional Krom clauses.
For a clause $\gcl$ we define three corresponding sets of propositional
Krom clauses:
\begin{itemize}
\item $\ren_1(\gcl) =
       \Set{\set{\skel(\gl)}}%
           {\gl\in\gcl,\,
            \exists\gv\colon \occ(\gv,\gl) > 1
           }
      $,
\item $\ren_2(\gcl) =
       \Set{\set{\skel(\gl), \skel(\glb)}}%
           {\gl,\glb\in\gcl,\,
            \gl\neq\glb,\,
            \var(\gl)\cap\var(\glb)\neq\emptyset
           }
      $,
\item $\ren_3(\gcl) =
       \Set{\set{\skel(\gl),\dual{\skel(\glb)}}}%
           {\gl,\glb\in\gcl,\,
            \exists\gv\colon\tmax(\gv,\gl)>\tmin(\gv,\glb)
           }
      $.
\end{itemize}
For a clause set $\gcs$, let $\ren(\gcs) = 
\bigcup_{\gcl\in\gcs}(\ren_1(\gcl)\cup\ren_2(\gcl)\cup\ren_3(\gcl))$.

$\ren_1$~encodes the fact that all non-linear literals have to be negative
in a clause belonging to a clause set in~\OCCiN.  More exactly, the sign of
a literal~$\gl$ where $\occ(\gv,\gl)> 1$ for some variable~$\gv$
has to be renamed if $\gl$ is positive and
must remain unchanged if $\gl$ is negative.
$\ren_2$~takes care of the fact that no two different
positive literals may share variables.
$\ren_3$~corresponds to condition~\ref{cond:OCCb}.

One can show that $\ren(\gcs)$ is satisfiable iff $\gcs$ is renamable to
\OCCiN. We even have:

\begin{proposition}
For every clause set $\gcs$, $\gr_\M(\gcs)$ is in \OCCiN\ iff $\M$ is a
 model of~$\ren(\gcs)$.
\end{proposition}

Again, renamability to \OCCiN\ can be decided in polynomial time
due to the fact that $\ren(\gcs)$ is a
polynomially bounded set of propositional Krom clauses.

\section{Testing for Renamability to~\PVD}\label{sec:PVD}
\begin{definition}
   A clause set,~$\gcs$, is in class~\PVD\ iff every clause $\gcl\in\gcs$
   and every variable~$\gv\in\var(\pc\gcl)$ satisfies the following condition:
   \begin{labenum}{PVD}
      \item \label{cond:PVDa} $\gv\in\var(\nc\gcl)$, and
      $\tmax(\gv,\pc\gcl)\leq\tmax(\gv,\nc\gcl)$.
   \end{labenum}
   $\gcs$ is renamable to~\PVD\ if there is a renaming,~$\gr$,
   such that $\gr(\gcs)$ is in~\PVD.
\end{definition}

In~\cite{FLTZ93} it is shown that the satisfiability of clause sets in~\PVD\
can be decided by hyper-resolution. Therefore the class of clause sets
which are renamable to~\PVD\ is decidable, too: just apply the decision
procedure to the renamed clause set.%
\footnote{In fact, \cite{FLTZ93}~defines \OCCiN\ and \PVD\ to include also
the clause sets renamable to these classes, and uses
semantic clash resolution with settings as decision procedure.
Finding a suitable setting corresponds to choosing an appropriate renaming.
Our view is justified by practice as it is easier to use a single fixed theorem
prover for hyper-resolution and to do the renaming as a pre-processing.}

We show that for a given clause set, condition~\ref{cond:PVDa}
can be encoded
by propositional clauses, which are satisfiable iff the clause set is renamable
to~\PVD. We start by illustrating the main idea by an example.
\begin{example}
   Consider the clause $\gcl=\set{P(f(x),y),\,Q(f(x),f(y)),\,\neg R(x,f(y))}$.
   $\gcl$~does not satisfy condition~\ref{cond:PVDa} because of variable~$x$
   which occurs at depth~$1$ in the positive part and only at depth~$0$ in the
   negative part. To satisfy~\ref{cond:PVDa}, at least one of
   the two literals containing $x$ at maximal depth,
   $P(f(x),y)$ and $Q(f(x),f(y))$, has to be `moved' to the negative part.
   This can be achieved by renaming either $P$ or $Q$ (or both).
   If we abbreviate the proposition \emph{`predicate symbol $P$ gets renamed'}
   by just $P$ (interpreting the predicate symbol as a propositional variable),
   the renaming requirement for~$x$ reads $P\lor Q$, or $\set{P,Q}$ in set
   notation. A similar requirement can be stated for~$y$:
   either $Q$ has to be renamed (making $Q(f(x),f(y))$ negative),
   or $R$ is \emph{not} renamed
   (retaining the negative literal $\neg R(x,f(y))$), which
   corresponds to $\set{Q,\neg R}$.

   If we denote an interpretation by the set of all atoms true in it,
   the models of the clause set
   \( \set{\set{P,Q},\,\set{Q,\neg R}} \)
   are given by $\set{P}$, $\set{Q}$, $\set{P,Q}$, $\set{Q,R}$, and
   $\set{P,Q,R}$.
   Each model describes a renaming which yields a clause in~\PVD\
   when applied to~$\gcl$. E.g., the model $\set{Q,R}$ corresponds
   to renaming $Q$~and~$R$ and leaving $P$ unchanged.

   In general the clause set consists of several clauses.
   To encode all restrictions we have to construct one propositional
   clause per variable and per clause.
\end{example}

For a variable~$\gv$ occurring in a clause~$\gcl$,
let $\clx\gcl\gv$ be the set of all literals in~$\gcl$ containing an
occurrence of~$\gv$ of maximal depth, i.e.,
$\clx\gcl\gv=\Set{\gl\in\gcl}{\gv\in\var(\gl),\,\tmax(\gv,\gl)=\tmax(\gv,\gcl)}$.
For a clause set $\gcs$, let $\ren(\gcs) = 
\Set{\skel(\clx\gcl\gv)}{\gcl\in\gcs,\,\gv\in\var(\gcl)}$.
\begin{proposition}\label{prop:PVD}
For every clause set $\gcs$, $\gr_\M(\gcs)$ is in \PVD\ iff $\M$ is a
 model of~$\ren(\gcs)$.
\end{proposition}

To determine the complexity of the renamability problem for~\PVD,
observe that the length of $\ren(\gcs)$ is polynomially
related to the length of~$\gcs$. By proposition~\ref{prop:PVD}, the renamability
problem reduces to the satisfiability problem for propositional CNFs.
On the other hand, the satisfiability problem can also be reduced to
the renamability problem for~\PVD:
given an arbitrary propositional CNF, interpret
each propositional variable as a unary predicate symbol and add
some dummy variable~$x$ as argument. The resulting clause set is renamable
to~\PVD\ iff the CNF is satisfiable. Therefore both problems are equivalent,
and we conclude that the renamability problem for~\PVD\ is NP-complete.

\section{Extracting Suitable Renamings}\label{sec:renaming}

In the last three sections we showed that the requirements for a clause set
to be renamable to Horn, \OCCiN, or \PVD, can be encoded as propositional
clauses. The clause set is renamable iff the propositional clauses are
satisfiable. Moreover, the propositional models are exactly the admissible
renamings.

One way to test the satisfiability of a propositional clause set~$\gcs$
is to saturate $\gcs$ under hyper-resolution; call the saturated set
$\hyper(\gcs)$. $\gcs$~is satisfiable iff $\hyper(\gcs)$ does not contain
the empty clause; in this case $\hyper(\gcs)$ can be used to find all models
of~$\gcs$.

Suppose for the moment that $\gcs$ itself is Horn. Then $\hyper(\gcs)$
is just the set of (singleton sets of) atoms
that are true in any model of~$\gcs$ (plus, of course, the initial negative
clauses). 
Thus, $\hyper(\gcs)$ is a representation of a model of~$\gcs$.

If $\gcs$ is not Horn we can still compute such a representation by 
replacing some non-singleton hyper-resolvent in $\hyper(\gcs)$ by one of its
literals and saturating the resulting clause set again. By iterating this
process of splitting and saturating we obtain an `atomic representation'
of a model of~$\gcs$. This procedure for building models
is described for general first order clauses in detail in~\cite{%
FermuellerLeitsch93,FermLeit98JIGPL}.
If $\gcs$ is Krom and propositional as in sections~\ref{sec:Horn}
and~\ref{sec:OCC1N} there
are only polynomially many different (hyper-)resolvents of~$\gcs$,
and the model building procedure terminates in polynomial time.

\section*{Acknowledgments}

We would like to thank the referees for drawing our attention to relevant
literature as well as for correcting some important typos.

\bibliography{gernot,salzer,renaming2Horn}
\end{document}